# AMHRP: Adaptive Multi-Hop Routing Protocol to Improve Network Lifetime for Multi-Hop Wireless Body Area Network

Muhammad Mateen Yaqoob [1], Kulsoom Fatima [2], Shahab Shamshirband [3,4*], Amir Mosavi [5,6] and Waqar Khurshid [7]

Department of Computer Science, COMSATS University Islamabad, Abbottabad Campus

[1] COMSATS University Islamabad, Abbottabad Campus. mateen@cuiatd.edu.pk

[2] University of Lahore, Sargodha Campus kulsoomfatima1988@gmail.com

[3] Department for Management of Science and Technology Development, Ton Duc Thang University, Ho Chi Minh City, Vietnam

[4] Faculty of Information Technology, Ton Duc Thang University, Ho Chi Minh City, Vietnam: Shahaboddin.shamshirband@tdtu.edu.vn

[5] School of the Built Environment, Oxford Brookes University, Oxford OX3 0BP, UK: a.mosavi@brookes.ac.uk

[6] Institute of Automation, Kando Kalman Faculty of Electrical Engineering, Obuda University, Budapest-1034, Hungary: amir.mosavi@kvk.uni-obuda.hu

[7] Waqar Khurshid COMSATS University Islamabad, Abbottabad Campus. waqarkhurshid@cuiatd.edu.pk

## Abstract

A Wireless Body Area Sensor Network (WBASN) is combination of numerous sense nodes, positioned onto/close or inside a person body. Wireless Body Area Sensor Networks (WBASN) is a developing automation trend that exploits wireless sensor nodes to put instantaneous wearable well-being of ill person to improve individual's existence. The sensor nodes might be used outwardly to observe abundant health parameters (like heart activity, blood pressure and cholesterol) of an ill person at a vital site within hospital. Hence the goal of WBASN is much crucial, enhancing the lifetime of nodes is compulsory to sustain many issues such as utility and efficiency. It is essential to evaluate time that when the first node will die it we want to refresh or change the battery reason is that loss of crucial information is not tolerable. The lifetime is termed as the time interval when a first node dies out due to battery exhaustion.  In our proposed protocol life time of a network is the main concern as well other protocol related issues such as throughput, path loss, and residual energy. Bio-sensors are used for deployment on human body. Poisson distribution and equilibrium model techniques have been used for attaining the required results. Multi-hop network topology and random network node deployment used to achieve minimum energy consumption and longer network lifetime.

## Keywords

## 1. Introduction

Wireless Sensor Network (WSN) is combination of "sensors nodes" that might be limited in number, cents either thousand in statistic. In this design a single sensor node associated with more sensor node or with various attached sensor nodes. Wireless Body Area Network (WBAN) is new emerging sub-category of WSN. WBAN is a combination of independent nodes (like sensors, actuators) that are placed in clothes, implanted on body or beneath the skin of a human. A major function of WBAN is health monitoring. Many diverse topologies are available in the network as star topology to advance multi-hop topologies. Network life is contemplated the time when first node or group of sensor nodes expires due to energy exhaustion.

Wireless Sensor Network (WSN) is a set of small sensor nodes that positioned in area to individually observe substantial fitness. WSNs calculate high amount of physical conditions as pressure, temperature, sound, etc. This sensed data than transfers to the base data (BS) which is gathered by sensor nodes. The applications for Military give arise in improvement and progress in the field of wireless sensor networks. WSN are bidirectional nodes that also regulate the actions of sensor nodes. Now a day, WSN are set up in numerous industrialized operations to supervise mechanical control, technical operations, and supervise machine condition. The WSN is combination of "Sensors nodes" that might be limited in number, cents either thousand in statistic. In this design a single sensor node associated with more sensor node or various attached sensor nodes.

Sensor nodes composite with different parts to perform different operations like to control the activities of node a microprocessor/microcontroller is attached, and for communication a radio transceiver is part of it, and an electronic circuitry to provide an interface to sensors for utilization of power sources. These sensors normally use Batteries for power source, or energy may be collected from some existing point of supply. The magnitude of node may have like shoe box or little sensor as dust particle, it may differ related to specific application. Same as this sensor nodes may have different cost according to sizes. It may vary from few to hundreds of dollars as nodes have compound circuitry and advanced structures. Many different topologies are available in the network as star topology to advance multi-hop topologies [1].

A wireless sensor network (WSN) has been discussed as dispersed network having scattered and independent detecting station. Every single sensing station is said to a sensor node. A sensor node comprise of a micro-computer, an energy resource (mostly a battery) and subject to the utilization field also use any sensor(s) and a transmitter, Few smart sensors prepared with electromechanical device, which is utilized to regulate the distinctive parts of the established structure, known as actuator [2].After formation of a network, sensor nodes compute, and collect statistics of any activity out of surroundings and report sensed data to base station in multi-hop fashion. Now a day Wireless Sensor Networks (WSNs) are being used to supervise different parameters in diverse applications from environment observation, to habitant monitoring, [3] war-field, farming and intelligent homes. These wireless sensors are scattered within sensing area to keep care full check on field. WBAN is a new emerging sub-category of WSN. Major function of WBAN is well-being monitoring. WBAN is a combination of independent nodes (like sensors, actuators) that should be placed in costume, implanted out of body or beneath epidermis of a human. Whole human body is contemplated a network and all nodes communicate through a wireless communication channel. Implementation of node network is done in the star or multi-hop topological manner [4].

Patient nursing is developing as a significant function of implanted sensors network. Many wireless sensors are embedded inside body or out-of-body of the patient. The collection of these miniature sensors creates a Wireless Body Area Sensor Networks (WBASNs). WBASNs can continuously monitor physiologic situation of person under supervision, and in case of emergency can deliver instantaneous response. In WBASN a patient is regularly observed, and in crucial condition a prompt action is requisite. These implanted sensors can assemble the physiologic information and then send to medical specialist in a hospital via Metropolitan Area Network (MAN) or Local Area Network (LAN).

At hospital, received information is examined carefully and on these base decisions is taken about victim's condition. WBASNs are used for health care and its applications are also outside the range of healthcare [5]. Wireless sensors are placed on human body or embedded within body to monitor critical signs of life such as blood circulation (BP), body temperature, heart activity, cholesterol level etc. Use of WBAN technology to monitor health parameters significantly declines expenditures of patient in hospital.

According to the reports of Department of Economics and Social Affairs of United Nation Secretariat [6] people of age 65 and over, which account for almost 15 percent of world population, will nearly double and became 761 million by the 2025. This fact reveals that by the half of century the health care will converted to a main concern. As the elderly aged folks are much susceptible to many fitness related diseases and affairs, so they have need of instant medical care and checkup, which results in high health care budget [7,8]. These facts and figures claim major modification and proactive management which focus on the avoidance, avoidance and early meditation of multiple infections [9].

With the assistance of WBAN technology, patients are observed at domestic level extensive duration. Sensors uninterruptedly feel data and advancing to health superintendent. WBANs have issue that sensor node are functioned with inadequate energy origin. It is requisite to employ least power for transferring information out of sensor nodes towards sink. The foremost complications for WBAN, is to recharge the batteries. A competent routing protocol required to overwhelm the issue of invigorating batteries. Numerous powers effectual routing protocols has been suggested for WSN technology [10], [11], [12].

However, WSNs and WBANs may have diverse manners, applications and function in dissimilar environment. The sensor nodes of WBASNs have diverse power level thus produce distinct sort of information. Whereas, sensor nodes of wireless sensor network (WSNs) have almost equal level of data rate and similar energy level. It's much difficult to implement WSN routing protocols for WBAN. Consequently, power proficient routing protocol for WBAN has been requisite to observe patients for extensive duration.

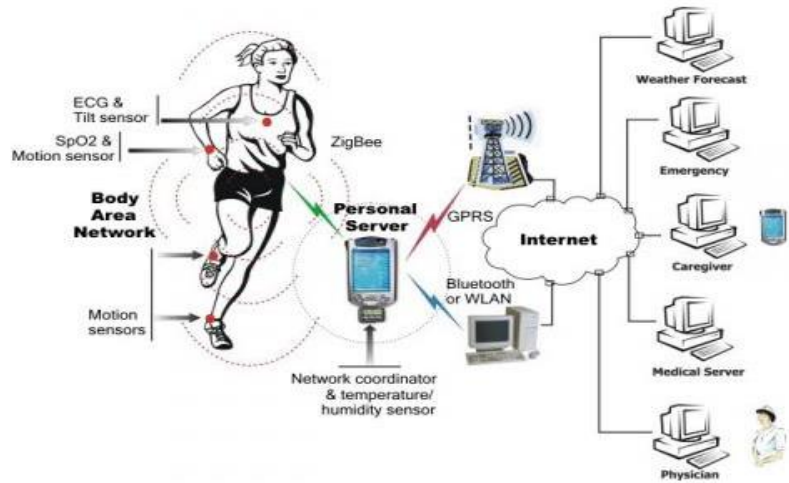

Figure 1: Wireless Body Area Network (WBAN) [4]

## 2. Architecture of WBAN

In WBAN, a node is denoted as an independent device having communication capability depending upon their application and functionality, and part in network, BAN nodes are categorized into different groups. A WBAN sensor node basically comprises on subsequent elements;

    i-       Sensing Component
    ii-      Processing unit
    iii-     A/D Converter (Analogue to Digital)
    iv-     A Power Unit
    v-      A Radio Transceiver (acts as communication unit)
    vi-     Storage Unit (i.e. Memory)

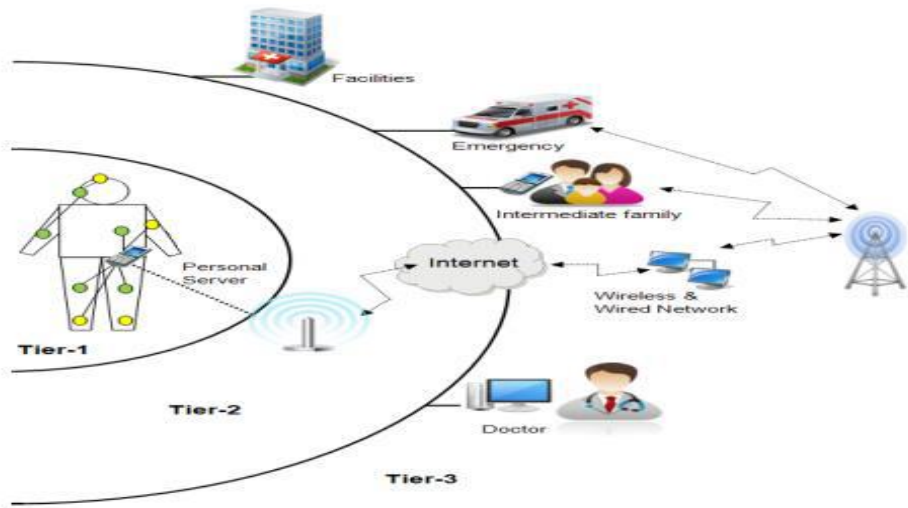

Figure 2: Structure of Wireless Body Sensor Network

In term of utility, following types of nodes available for WBAN.

(i)     *Sensor*; which measures definite parameters inside or within one's body and then after gathering and processing essential data furnish wireless reply to information.
(ii)    *Actuators;* once receives data from sensor it established link with the user [13].
(iii)   *Personal Device;* handles interaction with other users and gathers whole information which is received from sensor and actuators.

In the sense of implementation sensors nodes can be categorized into three different classes, which are embedded Node, outside body node and in-vivo node. One of them embedded in person body, 2 cm off from it or far away from body respectively [14, 15]. Based on their working in the WBAN sensor nodes can be classified into following categories;

a- *Coordinator*: It behaves like gateway (GW) to external cosmos or any other BAN. Coordinator itself merely a sensor node or it can gather data and work as relaying device. For advance processing coordinator node transfers sensed data via wireless access point (AP) to the supporting network. coordinator node also acts as an access point (AP) in few applications. [15]

b- *End Node*: End node is competent for accomplishment of their fixed requests.

c- *Routers*: A root node and child nodes have nodes act as an intermediate node through they broadcast information called routers.

Relying upon IEEE 802.15.6 standard BAN nodes did not supposed to function for either single hope or multi hop star topology in which central node can be placed on a body position like waist, chest etc. [16, 17]. With respect to communication architecture WBAN has been alienated in three dissimilar layers. For enabling an efficient and module-based system this communication structure covers many design issues. As shown in figure 2, in a centralized network architecture device of WBAN are distributed throughout the human body where definite position of any device is depend upon the relevance.

WBANs are sensor network which is not node dense dissimilar to common wireless sensor network. Due to this, there are not any redundant nodes which guarantee secure operation in case of letdowns in the network. Actuator or sensor nodes located, on definite positions on the human body [18]. Sensor node are supposed to use for recording physiological actions in periodic manner. Nodes are intended to record physiological activities of human in periodic custom and therefore data stream shows comparatively stable rates. These attributes make WBANs structure architecture little different from alternative sensor networks. Communication structure of WBAN perhaps build on either infrastructure-based communication or peer-to-peer communication [19], [20].

An infrastructure-based network consists of either on base station (BS) or on an Access Point (AP) which advancing traffic to the purposed receiver whereas in ad hoc networks peer to peer communication used. It is notable that in infrastructure base networks entire communications requisitely experience the BS or AP even if the sander and receiver both in the radio range of one another such as an AP or else BS is the supervisor of network.

In Figure 3 (a) an ad-hoc network demonstrated in which whole nodes can issue data circulation to more new nodes in reach of radio range depicted in Figure 1.3 (b) the infrastructure-based technique illustrated in [21]. Now a days, entire types of mobile telephone structure work within infrastructure approach and similar go for WLAN standard i.e. IEEE 802.11. Though, the second contribute basis for ad hoc mode too, for example that might utilized during a conference when conference members have desire to transfer documents [19]. Generalized structure of a WBAN should be distributed in following three significant communication levels as specify within [18]:

  i)   Layer-1 transmission (intra-WBAN)

  ii)  Layer-2 transmission (inter-WBAN)

  iii) Layer-3 transmission (beyond WBAN).

Wireless Sensor Network (WSN) is a set of small sensor nodes that positioned in a domain for individually observe personal fitness. WSNs calculate high amount of physical conditions as pressure, temperature, sound, etc. This sensed data than transfers to the base data (BS) which is gathered by sensor nodes.

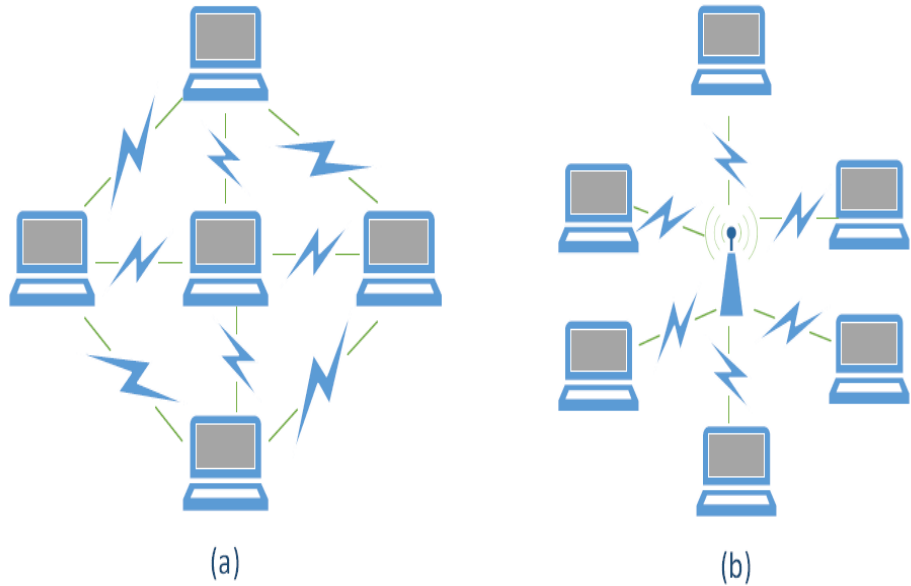

Figure 3:  a) Transmission in an Ad-hoc network    (b) Structure-type

The formation of WBASNs divided as subsequent layers

- In tier 1 the in vivo and embedded bio medical sensor nodes dispatch sensed data to either coordinator or ground station.

- In Inter-WBSNs, after collection and processing of desired information, coordinator or base station transmit received packets to sink(s).

- At this level the sink(s) transmit the gathered information to isolated medical center or any other object through established network. [22]

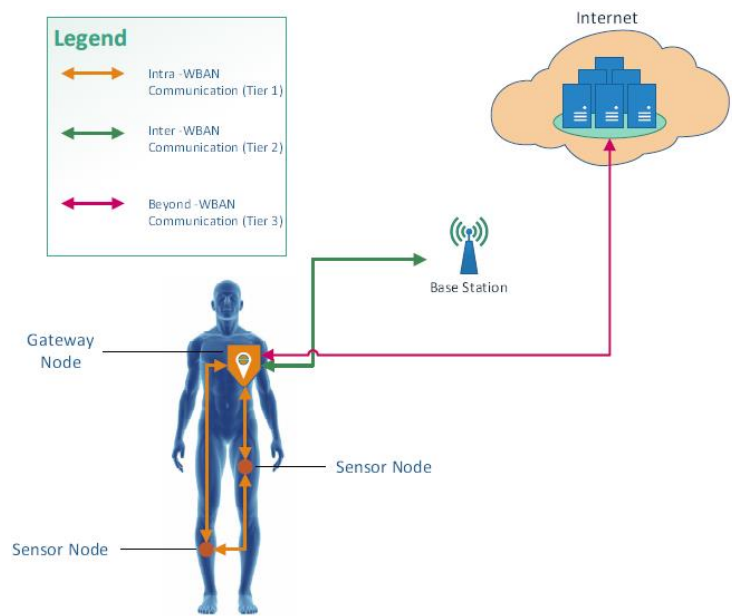

Figure 4: 3-Tier Communication in WBAN

### 3. WBAN Communication Standards

In WBAN sensor nodes may be implanted with in or into the person body. Sensor nodes communicate with nodes through specific limited scope wireless technology for intra- BAN communication. There are many communication standards are existing for WBAN. All types of microchip based wearable devices are also depends on these standards. From now on IEEE 802.15.6 (2012) has been purposed standard for WBAN but the marketplace for 802.15.6 -2012 dependent devices is under processing and premature. Now a days widely accepted standards include Bluetooth (802.15.1), MICS, ZigBee, Ultra wide band (UBV), IEEE 802 .15.6 which are briefly discussed below.

### i) Bluetooth (IEEE 802.15.1)

A communication standard that utilized limited-reach communication is Bluetooth (802.15.1) which is cultivated by Ericsson and his companion in 1998 in the Bluetooth Special Interest Group (SIG) [23].After this it have been standardized by the IEEE 802.15 ( WPAN Task Group (TG1)) also define the conventional term IEEE 802.15.1 [24] .In the early days of its development it is considered as a replacement for cable medium, but as time passes its use has been used in many diverse networking setups and applications. Many variants of Bluetooth are also available in market with the advancement in technology.

IEEE 802.15.1 communication standard typically used within the range of 10m, which is a short-range communication. Bluetooth Latency rate is minimum while the Bluetooth standard is having greater bandwidth. By using Bluetooth protocol, Bluetooth devices communicates non-line of sight (NLoS) with noble devices. Based on the extreme energy they permit to transmit; Bluetooth devices can be categorized into following classes.

Table 1: Categories of Power Classes of Bluetooth

| Power Class of Bluetooth | Maximum Output Power | Radio Range |
|---|---|---|
| Class –I | 100mW (20 dBm) | ~100 M |
| Class –II | 2.5 mW (04 dBm) | ~10 M |
| Class –III | 1mW ( zero dBm) | ~01 M |

Bluetooth standards generally have the data range limit of 3Mbps.Greater bandwidth boosts usage of Bluetooth standard UHC. This standard absorbs maximum energy, due to the reason usually this standard does not utilize UHC. Also, this standard does not suitable in favor of such kind of network where the issue of bandwidth and latency is critical.

### ii) MICS

For communication in medical applications a frequency band named Medical Implant Communication Service (MICS) is specifically designed. It is also support limited range wireless communication which ranging a person body. This standard is particularly used for near/on the body or inward the human body (but not only for mankind).

MICS is implant communication license band having the frequency rang in most of the countries from 402-405 MHz MICS gathers data from sensor nodes implanted within the body and scattered on body and, after this transmit this gathered data in multi-hop manners to the attached sink.

It utilizes minim power to transmit data for the safety to decrease Specific Absorption Rate (SAR) in contrast to Ultra-wide band (UWB). So, it minimizes the power consumption and also secure for sensitive human body tissues. It ensures Quality of Service (QoS), e.g. for giving emergency messaging. It also provides powerful security to those applications which carry sensitive information. [25]

### iii) ZigBee

IEEE 802.15.4 (ZigBee) is another commonly adopted communication standard for WBAN. It is specially used for communication devices which operate on low energy, for example sensor nodes. ZigBee acquire the power to regulate the complex operation associated with the communication issue. This communication standard hired with collision prevention technique. ZigBee consumes very limited energy during communication approximately 60mW. Therefore, data estimate of communication standard also limited to 250 kbps. ZigBee provides security due to its capability of data encryption which provides considerable protection during communication.

### iv) Ultra-Wide Band (UWB) IEEE 802.15.6

UWB is used for high data rate application because it is a high bandwidth communication standard. It has minimum emission power density due to this, it ensures long battery life. Whenever an application demands high bandwidth UWB is a finest choice. UWB is significantly excellent choice for use of emergency applications. It is implanted with Global Positioning System (GPS), which provides minimum routing path to access point or medical coordinator.

 Global Positioning System (GPS) will offer routes having low traffic rate that will make faster communication of critical emergency data to medical server. But it is not a good choice for wearable application because receiver used for UWB band is much complex.

### v) Wireless Medical Telemetry Services (WMTS)

A licensed band which is utilized medical telemetry system is Wireless Medical Telemetry Services (WMTS). Like MICS band WMTS also do not aid with high data rate applications. In a nutshell, 802.15.6 standard will define devoted signal structure of WBAN for future. The current standards for WBAN which are in use 802.15.4 with ext. 802.15.4a, and UWB with Physical Layer. During the year 2011, IEEE combine various other standards including their amendments and the original 802.15.4 within single standard labelled IEEE 802.14.4 -2011[26].

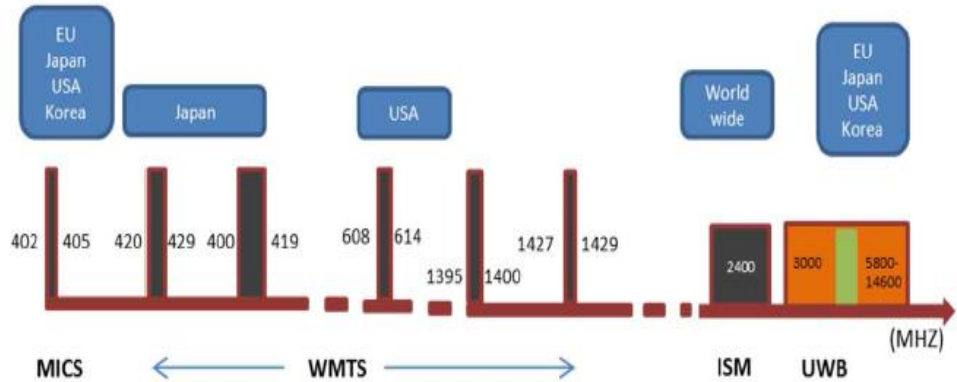

Figure 5. Frequency band for WBAN

## 4. Pitfalls and Challenges in WBAN

Design and establishment of an effective protocol for WBAN is a difficult task reason is that they have some distinctive requirements and attributes [28]. In given section we discuss the routing pitfalls and challenges.

### a) Inadequate Resources

WBAN have restricted energy resource as well having limited Radio frequency (RF) communication reach, inadequate storing capability, less computation abilities along with small bandwidth that may retain fluctuation because of noise and extra interferences [28].

### b) Privacy Violation

WBAN technology recently being used for the heath intensive care for the humans, though a few folk scrutinized that if WBAN technology cross the bounds of the protected medical usage it may cause danger to the liberty of humans.

### c) Diverse Environment

In a WBAN dissimilar type sensor nodes are essential for observing and sensing diverse fitness specification about individuals, which may have different storing ability, energy absorption and may also different in computation [29]. Therefore, this assorted essence of WBAN enforces some additional challenges.

### d) Radiation Consumption and Temperature rise issue

The two main causes of overheating node one is antenna energy consumption and other is power absorption of node electric circuit [30], in result it will disturb the temperature sensitive tissues of human anatomy [30] and can harm few tissues [31]. It is important for researchers to generate such protocols which retain secure human body tissues from the issues caused by temperature rise, radiation consumption and functions of embedded biomedical sensors.

e) **Constancy of Data**

The collective data of a patient sensed by a wireless sensor examined in a soothing way the important data about the patient exist on numerous nodes and then transferred to the medical server. For advance examination this data passes to the health specialist, if it's do not possible then the excellence of ill person health care and nursing process decline [32].

f) **Devices Structure**

The WBAN sensors are widely being used for the health care services thus the weight of theses nodes must be lesser, so they can simply be fixed or embed on human body. Sensors must have energy proficient because they must rout for quite a lot of years to observe patient. They must be accessible and reconfigurable. Moreover, the data must be saved to remote storing device so that medical practitioner can examine and make assessment via internet.

g) **Topological Separation**

Most of the time WBAN's network topologies must face difficulty of interruption or partitioning due to the limited range communication and body posture motion. Many researchers attempt to resolve the difficulty of partition and discontinuation with several methods. For instance, the researcher of [33] practice line of sight (LoS) & Non-Line of sight (NLoS) transmission, whereas author [34, 35] make usage of store and impudent routine for resolve the issue. Thus, the suggested protocols must conclude diverse topological variations.

h) **Path Loss**

Path attenuation termed as failure of power solidity for an electro-magnetic wave to broadcast throughout the wireless channel. Path Loss is the proportion of the power transferred to the received signals [36]. In the wireless transmission among the embedded sensor over the human body at which the path loss exponent may differ such as four to seven, and it extremely high if we make comparison with an independent space where it's only two. The scientists should take consideration the path loss issue while plotting any routing protocol for WBAN.

## 5. Literature Review

Wireless Body Area Sensors are being used to observe individual well-being with inadequate power means. Distinct power competent routing patterns are utilized to impudent data from person's sensors to health check server in hospitals. It has been significant that perceived data of patient consistently admitted towards medical practitioner for more examination.

Increasing heat of ingrained sensor nodes because of communication emission and electronic chips energy feeding can create influence on personal body. In [36], researchers employ thermal-aware routing to reduce the outcomes of growing temperature of planted sensor. Quwaider *et al.* [37] utilize Single-hop transmission that has enlarged communication scope of sensors for overwhelmed difficulty about division. A protocol named Environment Adaptive Routing (EAR) algorithm [38] describes dissimilar communication cost for heterogeneous WBASN gadgets. The suggested algorithm can be divided into the following three modules.

- Routing Table Constructor,
- Fault Detector,
- Path selector.

When coordinator node transfers a broadcast message to the entire attached node to build the routing table, individually node that obtains message runs the Routing table construction module to build its routing table. Though, the Single-hop communication and preemptive routing isn't appropriate option against WBASNs. Multi-hop communication is appropriate as ordinary packet deliverance and due to high transmission cost only Single-hop is utilized for crises maintenance. However, after a steady interval use of Hello messages at this point, consequences in immense power feeding.

The author of [39] enforces an algorithm named tree algorithm by prioritizing for WBASNs which can be used in health care and non- therapeutic domains. In [39] a ordering for crises situation for health care and non-health care WBANs application is presented. A peculiar medium is devoted for crises data deliverance whereas the ordinary data communication will be delayed till the fortunate transfer of crucial information. However, a devoted medium will result in the damage of existing resources. A collision settlement protocol, Tree algorithm is used to accomplish the objective. This algorithm permits the high precedence nodes to obtain the carrier earlier without competing small precedence nodes.

M-ATTEMPT [40] is a power proficient & temperature aware routing protocol for WBSNs. It decreases nodes heat in addition to this also minimize the lagging for the acute data utilizing diverse bio-medical sensor nodes. A network design with this arrangement, nodes having huge data rate are positioned at inferior movable region of the personal body whereas the sink node act as base station (BS) is positioned at the middle. For crucial or request driven information packets the sensor nodes of this scheme enhance their communication energy for advancing the information packets precisely to the sink node (single-hop).

Then again, for common data packets delivery multi-hop transmission technique is utilized for this purpose. The sensors which have common information packets cannot advance them as for as the sink (base station) accepted entire crucial or request motivated data packet. While in case of multi-hop communication, a path along minimum leap-count carefully choose either double or multiple paths are accessible. Whenever, multiple upcoming-hop adjacent nodes have similar leap-count then the adjacent node with minimum power expenditure toward sink will be chosen.

To restrained warmth, increase issue, M-ATTEMPT states a threshold value if any node lead overheated elsewhere its value, it breakdowns whole possible paths with the adjacent node. Though, as for as node's heat grasps to the threshold value once getting information packet, it again sends this packet to preceding node thus preceding node show it hotspot. Diverse M-ATTEMPT stages include:

(i) Initialization stage,

(ii) Routing stage,

(iv)    Organizing stage,

(v)     Information transmission segment

All the above mention stages are shown in diagram 2.1. At the opening stage entire sensors send out hello packet, whereas in second stage the routing phase paths with minimum leap-counts are chosen inside the existing paths established on above-mentioned approach. The sink node (BS) produces Time Division Multiple Access (TDMA) scheme concerning overall base nodes at scheduling stage, although throughout information transmission segment root nodes dispatch their information towards sink.

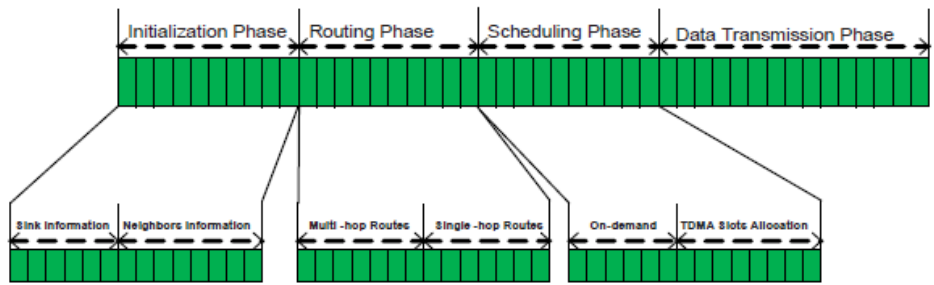

Figure 6: Sequences of Phases in each Round [40]

In [41] the writer intended to a cross-layered protocol, termed as wireless autonomous spanning tree protocol. This protocol distributes period spindle in slits within dispersed way, recognized WASP-Cycles. A WASP-Cycle can deliver circulation routing and medium access allocation applying the identical spanning tree mechanism that effects in minor energy feeding and greater throughput. The spanning tree is will be established spontaneously as depicted in Figure 2.2. This scheme is utilized to dispatch the information to the sink and integrates the media access control and routing.

In unique WASP scheme message is assign to each node which sends it to its child nodes instructing and them at what time they permitted to utilize the connection. Therefore, these dispatches permit stream of traffic and rise resource invitation from root of children that minimize the coordination operating expense. Based on, sink WASP-structure and the child node's demands the children reply to the structure by transmitting their specific WASP-pattern. It is essential to know that nodes are co-ordinate each other to escape fluctuations. For sink and child nodes, WASP-scheme messages are different, but they typically comprise of the following:

i)      Location of sender node

ii)     Slots already allocated to child of root node where they transmit their    WASP-structure

iii)    Quite time interval

iv)     Shipping accepted information to the sink

v)      Contention slit

vi)     Acknowledgment order.

Results show that WASP accomplishes almost 94 percent throughput, maximum packet deliverance proportion, little energy feeding and stable adjoining delay. Though, WASP does not acknowledge excellence of link, movability, weight and not endorse interactive connection. Overhead is also a weakness that may be minimizing with data accumulation procedures.

Latre *et al*. [42] proposed A Secure Low-Delay Protocol for Multi-hop Wireless Body Area Networks (CICADA) routing protocol which consists of a spanning tree structure similar to WASP [20] as illustrated in diagram 2.3.A cross-layered small power protocol CICAD used Time Division Multiple Access (TDMA) scheduling which is founded on grounds of multi-hop movable body area networks. In its Time Division Multiple Access (TDMA) protocol schedules the transportation for nodes.

Time Slits are allocated in scattered custom and identified extent of every cycle develops slot synchronism promising. In this protocol parents' nodes are accountable for informing their children when they will be interconnecting. The nodes close the root perform the way as forwarder nodes or parent nodes, these nodes gather information from their related children nodes and transmit to sink. Owing to additional traffic burden of children nodes upon root nodes causes parent nodes to exhaust their energy fast. This arrangement splits every cycle in to substitute-cycle: one is the controller sub cycle and second is data substitute cycle.

Every sub-cycle assigns the slits allowing to their individual structure: the information scheme and controller scheme, and throughout the control sub-cycle from the parent toward their child nodes both are forwarded. Respectively, control scheme demonstrates order, in that the child node can transmit their parents their owing control scheme, intensity of the tree and also extent of controller sub-cycle. The information sub-cycle begins after termination of the control sub-cycle. The extent of waiting time and the extent of data time are contents of information scheme. In data time, child node transmits data packet towards their root node. To save power, throughout the waiting period node can also go to sleep mode.

Every parent node builds a counter for its child nodes holding sum of slits they want to transfer information to their origin node and collection of slots mandatory to obtain the information dispatched from their child nodes.

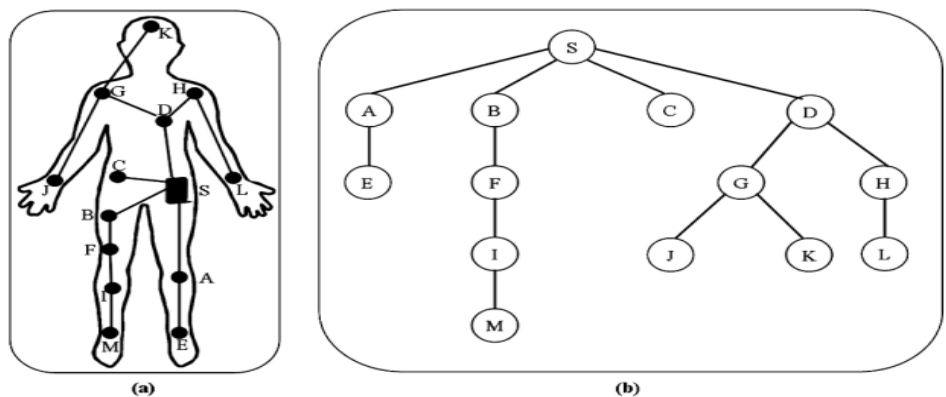

Figure 7: (a) On-body Network         (b) Summarize prospect of network [22]

Just before attaching the tree structure every data sub-cycle has a slit for fresh nodes. Every fresh child node has permitted to convey JOIN-REQUEST, dispatch in that slit afterwards listening the information structure of the coveted parent node.

In [43] the authors believed that WBASN occupy an important part to get the instantaneous and information by condensing the scope of energy feasting. WBASN includes of minute, limited power, and inconsequential sensors located in vivo or inside individual's body, to observe some uncertainty in frame organs and collect numerous biomedical constrains related to patient. In given research protocol termed Distance Aware Relaying Energy efficient (DARE) just observe ill personal by using multi-hop Body Sensor Networks (BSNs) is proposed. Presented protocol functioning via scrutinizes area of a hospital containing multiple patients briefly eight.

They experienced diverse topologies through putting the sink at dissimilar positions in ward, fabricating it fixed or else movable. In a hospital scenario seven sensors including Body Relay (BR), continuously data monitoring sensors (BS) and event-driven data (Threshold) monitoring sensor (BS) are attached to each patient. These sensor nodes measure different parameters related to patient's health which includes Electrocardiogram (ECG), pulse rate, heart rate, temperature level, glucose level, toxins level and motion. To decrease the energy exhaustion, sensors creates a link with sink by means of an on-body relay, fasten on trunk of each patient.

A relay attached on victim body retains greater energy possessions as distinguished to the sensors affix on body as; they perform amassment and relaying of data to the sink node. Hence, for heterogeneous networks; a relaying energy-efficient protocol for constantly observing patients is suggested. In the given scenarios, few sensors nodes constantly monitor patient's data whereas, some other nodes monitor data only when they notice a definite threshold stage. The protocol defines minimum power criterion for sensors to avoid damage to sensitive body organs of victim. Results of the research depict greater packet delivery ratio, longer Network life time and better stability period but it has high propagation delay.

In [44], the researcher recommended opportunistic routing protocol through bearing in mind stirring temperament of person's body. Network prototype utilize the protocol is much ordinary, whereas a bio-medical node affixes the trunk, gathers the data and dispatched it to the other node i.e. sink which is attach to the patient. Furthermore, a relay node located on wrist to simplify the connection among the sink node and bio-medical sensor. Throughout, the movement (pedestrian or running) of person body the carpus (which consists sink and relay nodes) moves onward and reverse causing duel types of promising connection establishment: (a) None Line of Sight (NLoS) connection—at time when wrist is behind backward of person's figure (b) Line of Sight (LoS) connection established when wrist is at frontal position of the body. None Line of Sight (NLoS) and Line of Sight (LoS) both connections are reflected to acquire similar likelihood, which approximately 0.5 ratio.

If sensor node has information packets and wants to dispatch to sink, it will direct Request to Send (RTS) packet, this only will be acknowledged by nodes at LoS. In case of RTS packet is recognized inside certain time, i.e., coordinator node in LoS with sensor, thus sensor node begins promoting information packets immediately to sink.

Furthermore, sensor node will transfer awaken call to relay node, if RTS packet not recognized within definite time interlude, i.e., the sink node is NLoS with sensor node. When relay node is prepared for connection, it notifies jointly to sensor node and sink for initialize transmission.

Subsequently, accepting all information packets successfully, sink node then dispatch Receive Acknowledge (RACK) packet towards nearest sensor node. In case, sensor node did not receive RACK packet, thus above-mentioned practice recurring up to prosperous connection.

Culpepper et al. [45] intended a data aggregating based protocol based designated as Hybrid Indirect Transmission (HIT). This protocol is established on single or more than one cluster, whither every cluster is competent with several multi-hop broadcasts. To reduce energy absorption and the network delay HIT utilize equal computation either in between-cluster or within-cluster connection. In this research, analysis oh HIT and HITm (HIT having multiple clusters) shows minimum network delay, increased network life time and greater energy efficiency.

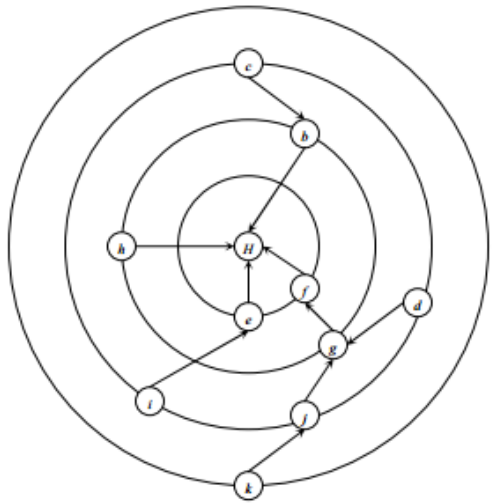

Figure 8: Cluster head selection in Hybrid Indirect Transmission (HIT) [45]

The detail procedure of HIT is as pursue: In the beginning point, sole or many cluster heads (CH) has been selected. Then throughout network cluster heads (CH) dispatch their position information. In adjacent step upriver and downriver connection of CH established. Therefore, numerous ways are structure inside a cluster to CH. Thus, every node computes its individual obstruction group.

The obstruction group of nodes $i$ consist record of nodes which have not been permitted to convey instantaneously among node $i$. further precisely, node $i$ jams node $j$ exclusively, where,

$$D(i\,; ui) > D(i; uj) \qquad (2.1)$$

In this equation upstream neighbor of node j is uj. At present, Time Division Multiple Access (TDMA) agenda calculated for individual node that permits extreme connection between nodes thru comparable broadcast. At the end, nodes coveys to their upriver acquaintance over TDMA agenda formerly allocated. Unluckily, HIT needs extra transmission power for thick networks and its requests as well as transmission ways. HIT do not deliberate dependability and also contradictory contact problems amongst demanded ways.

In [46] an improved version of Power-Efficient Gathering in Sensor Information Systems (PEGASIS) presented. This paper gives the idea of flexibility of sink enhanced the energy efficient PEGASIS-based protocol (IEEPB) for development of network duration concerning Wireless Sensor Networks (WSNs). Idea of sink flexibility, multiple-head and multiple-chain greatly influence for increasing the network duration of wireless sensor nodes.

The recommended idea of Mobile sink improved energy-efficient PEGASIS-based routing protocol (MIEEPB); is a multi-chain model which provide mobility of sink, to attain effective energy usage of wireless sensors. There is a need to enclose its movement within boundaries and the route of mobile sink should be static as the automated shifting of mobile sink is conduct by petrol or current. In this technique, the mobile sink moves along its route and stays at sojourn location for a sojourn time to give affirmation of comprehensive data collection.

An algorithm has developed for route of mobile sink and performing wide-range of experiments to evaluate the achievement of the advised approach. The outcomes of this research declare that proposed scheme is almost excellent and also performs exceptional than IEEPB in terms of network lifetime and stability period. It controls the ordinary power exhaustion in nodes, shrinks delay in data delivery by means of smaller chains.

Researcher in [47] suggested an algorithm, namely thermal aware routing algorithm (TARA) which intends towards minimize opportunity of temperature rise within embedded sensor network. Application environment of the implanted network is temperature sensitive. Due to the mass and plainness of the bio medical sensor node, the researcher of [48] supposed that there is not any sensor node within bio-medical node which can compute the warmth of sensor node.

- At initial stage each node notices the movements of its adjacent node, evaluates their communication emission and power utilization by counting the acknowledged/broadcasted packets. Depending upon these perceptions each connected node evaluates the modification in temperature of its adjacent nodes. Hotspots are those nodes which have the temperature apart from the threshold amount.

- In second stage; the routing phase, those data packets which do not have hotspot node as terminal node are routed over nodes apart from the hotspot nodes applying retraction strategy. Scheme of retraction strategy is that nodes which have hotspot as their adjacent node post back their data packet to the operator node to discover substitute routes to advancing the data packet, which presented in diagram 9.

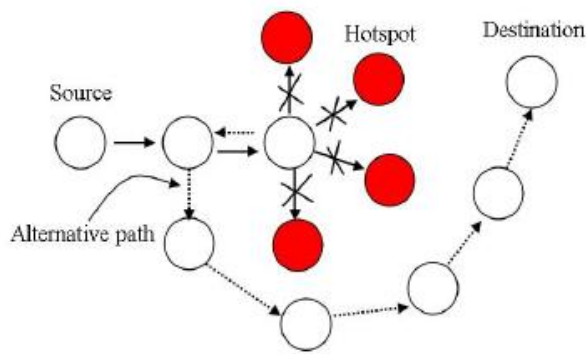

Figure 9: An illustration of thermal aware routing algorithm (TARA) [47]

The routing strategy describes that the data packets which have hotspot sensor as terminal node, absorbed till over-heating issue of destination node decline to a definite value.

The system model of thermal aware routing algorithm (TARA), specify that it is supposed that insulator and infusion characteristics of human organs has been acknowledged. Moreover, bio-medical sensor objectively embedded to individual. Embedded bio-medical sensors constantly impression information and then deliver towards gateway (GW)/coordinator that liable being combining information and forwarding towards base station (BS) placed further the human body. TARA measures the temperature and Specific absorption Rate (SAR) of every sensor by applying the Finite Difference Time Domain (FDTD) and Penne's Bio-heat Equation mechanisms.

In [48] a data collecting algorithm anybody utilize clustering for minimizing amount of straight communication to remote BS (base station). Its model founded upon the working model of LEACH protocol. LEACH picks a CH randomly in fixed time period for sake of spreading energy dissipation. The data dispatched to base station via cluster head, accumulates entire data then forward. LEACH have a problem is about the assumption that all the network nodes are not outside the sending spectrum of the BS. Anybody resolves this problematic issue by establishing a backing support network for cluster head and altering the way of cluster head selection. Although not examined the reliability of the network nor completely scrutinized the energy efficiency issue.

Aforementioned protocols may categorize in different types of routing protocols. The researchers divided them in to heat aware, postural movement based, cluster based, cross layered based routing protocols. M-ATTEMP [40] is heat aware routing protocol.

Contradicted to concept of multi-hop communication the temperature aware routing protocol M-ATTEMPT minimizes the energy consumption and temperature leap. This protocol has the increased power exhaustion, packet deliverance proportion and warmth leap. This protocol has the increased packet delivery ratio. A cross layered protocol for WBANs is WASP [41] that enhance packet delivery ratio. Other benefits of WASP include, it minimizes the energy absorption of nodes and in addition to this also reduces the end-to-end delay. Like other protocols it also has some drawbacks, which has its poor link established quality, and it does not upkeep shared communication. An improve variation of WASP is CICADA [42], also a cross layered protocol.

CICADA has sufficient sleep time as compare to WASP; consequently, it declines energy consumption for sensor nodes. DARE [43] consumes less energy due to sink nodes attach on chest of each patient. This protocol presents five different scenarios for placements of node on patient's body, which results in longer network life time and greater stability period. But these scenarios make DARE owning high propagation delay due to its distance with sink.

The opportunistic protocol [44] is a postural movement-based routing protocol, having the whole energy expenditure amongst multi-hop and single hop communication. It has equal level of energy expenditure for sensor and relay node; i.e. 50 % for relay and 50 % for sensor nodes. In opportunistic protocol a separable energy level is not measured; however, the amount of nodes rise in network, circulation load on the relay node will automatically increases.

HIT [45] absorbs much power for thick or large network. However, to establish a direct link for lesser number of sensor nodes in a network, HIT performs well for minimum energy consumption. Additionally, it is beneficial for data gathering due to its less network delay. HIT does not bother packet delivery latency and shows reduce packet delivery delay and energy absorption.

MIEEPB [46] is an extended version of refined energy efficient PEGASIS-based protocol (IEEPB), with idea of mobile sink. MIEEPB has provided an optimal approach and enhance the network lifetime as compared to previous version. In [49] authors are evaluating the performance of IEEE 802.15.4 for multi-hop communication within low power sensor network by varying data rates and frequencies. Authors have done this by under variable frequencies and data rates.

Table 2: Comparison of different Routing Protocols

| Characteristics | Protocols | | | | | | |
|---|---|---|---|---|---|---|---|
| | M-ATTEMPT | WASP | CICADA | DARE | Opprt. | HIT | MIEEPB |
| **Network Lifetime** | Normal | Great | Great | Great | Great | Normal | Much High |
| **PDR** | Great | 100% | --- | Up to 80 % | 100% | --- | --- |
| **Energy cons.** | Small | Small | Very Small | Small | Very Low | Small | Normal |
| **Delay** | Small | Small | Very Small | High | Small | Small | Small |
| **Mobility** | Yes | --- | | Yes | Yes | --- | Yes |

As the increase of possible hazard of bio-terrorism, there is a huge space for the tool that can be rapidly, attentively, and accurately distinguish bio agents spread in the surroundings. Biosensors can be function as economical and greatly effective gadget for detection of bio agents and furthermore, may also be used in everyday applications. Different terms and definitions widely used about biosensor depending upon the specific use and their applications. They are also termed as bio-chips, glucometers, immune sensors, chemical canaries and bio-computers. One of the two most commonly mentioned definition about biosensor quoted by S.P.J Higson,

"A biosensor is a chemical sensing device in which a biologically derived recognition entity is coupled to a transducer, to allow the quantitative development of some complex bio-chemical parameters."

And secondly D.M. Frazer said about biosensor that,

"A biosensor is an analytical device incorporating a deliberate and intimate combination of a specific biological element (that creates a acknowledge event) and a physical element (that transduces the acknowledge event)." [50]

As the label biosensor suggests, device is mixture of two different chunks, one is the bio-element and the other part is a sensor element. The elementary perception about biosensor's process demonstrated with figure 10.

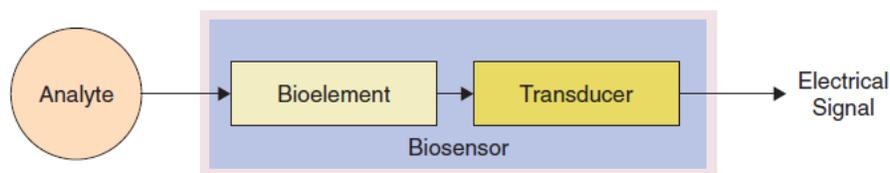

Figure 10: A graphic illustration of Biosensor

A bio element like an enzyme identifies analyte and the definite sensor element transduces conversion in biomolecule within the form of an electric signal.

Bio element much definite to concerning analyte, for which it is very delicate. The reason is that it does not identify any other active analytes. The biosensors may have multiple types, but it depends on their undergoing transducing mechanism. They have been divided into subsequent types: i.e. [51]

  i.    Resonant Biosensor
  ii.   Thermal detection Biosensors
  iii.  Ion-sensitive field-effect transistor biosensor (ISEFT)
  iv.   Optical detection Biosensor
  v.    Electro-chemical Biosensor

The electrochemical type biosensors, founded on the stricture measured, so they additionally categorized into further three types [52]. Such as

- Conducti metric
- Ampero metric
- Potentiometric.

## 6. AMHRP: Adaptive Multi-hop Routing Protocol

Body sensor network is gaining importance day by day. Patients are gaining benefits from this kind of technology improvements. This particular field has merged various disciplines like medical science, data communication, computing. All these areas are now working together in order to establish best possible mechanism that can achieve longer lifetime network. Our purpose routing mechanism has improved the lifetime and minimizes the energy constraints. Moreover, also has provided the recharging mechanism for bio sensor. Our research has also focus on developing the used dependent routing mechanism. Most of the researchers have purposed the communication of bio sensor through a sink but what happens when sink stops to respond?

Networks have long been recognized as having a central role in biological sciences. They are the natural underlying structures for the description of a wide array of biological processes across scales varying from molecular processes to species interactions. Especially at smaller scales, many genes and proteins do not have a working their own behalf; somewhat they acquire a specific role through the complicated mesh of connections with other proteins and genes. In recent years this viewpoint, mainly nurtured by the current profusion of high-throughput experimentations and availability of entire genome sequences and gene co-expression designs, has ran to a stream of activities concentrating on the construction of biological networks.

The abundance of large-scale data sets on biological networks has revealed that their topological properties in many cases depart considerably from the random homogeneous paradigm. This evidence has spurred intense research activity aimed at understanding the origin of these properties as well as their biological relevance. The problem amounts to linking structure and function, in most cases, by understanding the interplay of topology and dynamical processes defined on the network. Empirical observations of heterogeneities have also revamped several areas and landmark problems such as Boolean network models and the issue of stability and complexity in ecosystems. While concepts and methods of complex network analysis are nowadays standard tools in network biology, it is clear that a discussion of their relevance and roles has to be critically examined by taking into account the specific nature of the biological problem.

Even sensors are in better shape but due to single point of failure we will not be able transceiver the data. Secondly, mostly researchers claimed to have static sensors and distance from the sink to sensor will remain constant. So, distance matrix will be of no use for these kinds of solutions. Cell functioning is the consequence of a complex network of interactions between constituents such as DNA, RNA, proteins, and other molecules. A wealth of information on the cell is encoded in the DNA which includes both genes and non-coding sequences. The genes preside over the production of proteins and other molecules, which themselves interact with the genes on several levels. For instance, transcription factors, proteins that activate or inhibit the transcription of genes into mRNA, are the product of genes that may be active or not. Ultimately, genes regulate each other via a complex interaction pattern forming the genetic regulatory network. Proteins in their turn perform functions by forming interacting complexes.

The map of physical interactions among proteins defines the protein–protein interaction network. Cell metabolism can also be thought of as a network whose fluxes are regulated by enzymes catalyzing the metabolic reactions. In general, all of these interaction networks are connected through a cascade of biochemical reactions when stimulated by a change in the cell environment, activating an appropriate transcriptional regulation that, in its turn, triggers the metabolic reaction enabling the cell to survive the change. Researchers have long been constrained to focus on individual cellular constituents and their functions, but the growth of high-throughput data grouping techniques now provides the possibility of simultaneous investigation of many components and their interactions. Microarrays, protein chips, and yeast two-hybrid screens are some of the techniques that allow the gathering of data on gene, protein, and molecule interactions. These data can be used to obtain a system description usually mapped into interaction networks, which in turn form the network of networks that regulates cell life.

These experiments are by construction error-prone, and many false positive and negative signals are usually found in the resulting data sets. This evidence has stimulated lively debate on how far these data sets can be considered reliable, and it is clear that continuous checking with specific high-confidence experiments is needed. On the other hand, for the first time it is possible to gather global information on cell functioning that challenges the research community to develop a systematic program to map and understand the many cell networks.

A first relevant result derived from the systematic analysis of a variety of cellular networks is that their topology is far from random, homogeneous graphs. On the contrary, the degree distribution of most cellular networks is skewed and heavy-tailed. In addition, it is generally accepted that these networks are fragmented into sets of diverse molecules or modules, everyone being accountable for dissimilar cellular functions. One of the first pieces of evidence comes from the analysis of protein interaction networks (PIN) where nodes are single proteins and a link among them represents the possibility of binding interactions. In all cases the topological analysis shows that the PINs have a giant connected component with small-world properties and a degree distribution which strongly differs from the Poisson distribution of homogeneous random graphs. The network is characterized by sensors and presents non-trivial correlations, as measured by a slightly disassortative behavior of the degree–degree correlations, and by the clustering coefficient.

Further evidence for the complexity of biological networks is found in transcription regulatory networks, in which the nodes correspond to genes and transcription factors and directed links to transcriptional interactions. In this case the corresponding networks again exhibit heavy-tailed out-degree distributions while an exponential behavior is found for the in-degree distribution. These properties provide valuable information on the mechanisms presiding over gene regulation processes. The broad out-degree distribution signals that there is an appreciable probability that some transcription factors interact with many genes. In contrast, each gene is regulated only by a small number of transcriptional factors as indicated by the exponential in-degree distribution.

We have purposed to have multipurpose bio sensors implantations rather than deploying a specific sensor. Although our suggested solution can also accommodate those sensors as well but with limitations. We have suggested a mechanism in which sensors are deployed in generic way i.e. 19 sensors across the body. If Physician suggests some specific sensors, then they will be addition to that 18 sensor. Our proposed routing protocol is adaptive in nature and capable of working in critical and normal situations. In critical situations it might happens that some specific data set required more frequently answer as compare to normal situation. In modern bio sensor technology computational and analysis engines are also embedded. We have tried to take advantage of these computational units and configured the routing mechanism upon these nodes. Before discussing the routing protocol, it is necessary to understand the network lifetime.

Network lifetime means the average life of implanted in the body and how longer they are capable of sending correct information. Most importantly human life is more important than the network lifetime and in emergency situations where data is shared often shorter intervals and it might cause early exhaustion of network life. Chief responsibility of our routing algorithm is to conserve maximum energy in normal situation and make use of resources in critical situation. Moreover, recharging phenomena of bio sensor is also purposed. Secondary ambition of routing protocol is to ensure correctness of sender data because false information may have to casualty.

- Firstly, we explain the working of protocol under normal situation, then for emergency situation.
- Secondly (Latterly), we have proposed the mechanism recharging of bio sensors and methodology for ensuring data correctness.
- Thirdly, purposed works specifies the integration of bio/body sensor with WSN from where information can be shared with outside world.

Multipurpose sensors capable of recording various readings at a time more over computing can also be performed at their end. Physicians suggest values and thresholds already been provided to sensors. It is assumed that generic topology would remain same and types of sensors can be replaced, if there is a need of implanting specific (single purpose) sensors then it can also be implanted in adding to generic multipurpose sensors. The key is to minimize the energy consumption in favor to enhance network lifetime.

We have attached weights for different types of packets. The detail of weights is given below,
1. Energy of self-computation is Xs.
2. Energy of sending (destined node) is Xd.
3. Energy of sending to WSN is Xw.

Where $100 \times Xd = Xw$

4. Energy of forwarding pack is Xf and its very small
5. Energy of exchanging control packet is
$X_w$ , it is higher then

$Xf$ & fewer then $Xd$

Where X is the aggregate energy related with each sensor individually and it keeps of declining unit X, when it will recharge and reaches the energy nearer to Xt, a threshold is attained. The life of each sensor in the network is specified here as:

$$Xt = n_1\ Xs + n_2 Xd + n_3 Xw + n_4 Xf + n_5 Xc \qquad (7.1)$$

Where n1, n2… n5 labeled as the frequencies, how many numbers of time that individual component used. We can compute the full life of the network with the assistance of distributed function. A node dies when there is $t \rightarrow T$. Our suggested routing protocol is adaptive in nature and will perform in accordance with the patient's condition and it may differ from one patient to another.

Assume that multipurpose sensor nodes implanted in human body sensed the data and then passes to the sink for further monitoring. Sink is attached in the center of the body, and liable for data aggregation and then forwarding information to the external medical server. Communication flow rely form sensor to sink or sensor to sensor and to sink, but communication outside the sink is not considered in this protocol. Forwarded node is selected on the basis of maximum residual energy or has minimum distance to sink. Multi-hop communication topology functioned relay nodes, which minimize the communication distance, consequently decrease the energy consumption on nodes and also increase the life time of a network. In continuous flow of data consumes lot of battery power and in result nodes die early. Type of reporting data in AMHRP is event driven, as it takes reading of data from periodic intervals or event is generated by advice of a medical specialist (doctor). For such type of extensive network, time division is a cheapest solution for collision avoidance. AMHRP can be a time driven protocol as each sensor have different time slot for data transmission.

When an event is takes place, it specifies the occurrence of threshold value, but it vary from patient to patient. Normal adult person heart rate ranges from 60-100bpm, when it increases to 100bpm or decrease from 60 bpm, it will indicate a threshold for ECG sensor. The lower temperature value set to be 36.5-37.5° C (97.7 – 99.5°F), and the higher life-threatening limit foe medical energy is 40°C (104° F) or above. The diabetic condition for critical level of glucose are minimum 110mg/dL to maximum 125mg/dL. Normal blood pressure a is 120/80 millimeters of mercury (mm Hg) or less and 140/90 mmHg is considered as high blood pressure. Despite this all other sensors like insulin, pressure, positioning, DNA protein, EMG, respiration, toxin, lactic acid, tilt, pH, SpO2 starts transmitting data when a maxim or minimum limit indicating the alarming condition. All the above-mentioned sensor checks biological behaviors of a patient according to the prescription of a medical specialist. It may vary from patient to patient regarding to her/his disease and condition of patient. Sensing of most significant feature in normal situation may occur multiple times or day or during week or may be during month as per advice of a physician during assigned time slot. Most significant features which required to be observed are

1. ECG                          (Once in a week)
2. Blood Pressure (B.P)         after every 3 hrs.

3. Glucose                  3 times a day (if diabetic)
4. Insulin                  once in a day
5. EMG                      Once in Month
6. Body Temperature         3 times a day
7. Enzyme Test,SpO2         once in a month
8. DNA protein, respiration, toxin, lactic acid, tilt( as per advised by the doctor )

Besides this some supplementary statistics also essential for examination such as walking Indicators, anxiety, tension, every day temperature, food analysis and stress etc. These type of parameters assists the surgeon for the recommendation of healthier treatment. The multi-functional sensor would work in accord with stochastic manner and will be capable to distribute power between body sensors and also specify the scheduling for ensuring enhanced and improved lifetime for imbedded sensor nodes. Therefore, it is essential to probability space related with randomly chosen sensor and probability dispensation is computed. Upon the described distribution scheduling is developed and judgment concerning to a sensor node to be a sender, forwarder or a cluster head is determined and confirming enhanced life time for that specific instance.

If for any objective a sensor node is designate at any time with average of $\lambda$ events happening at that point distribution of frequency of occurring events, S is

$$P(k) = e^{-\lambda} \lambda^k / k! \qquad (7.2)$$

We obligate to make use of Poisson distribution and ensemble average of random sensor is essentially the average of embedded sensor of all accomplishments. WBSN comprise multi-hop, multiple frequency bands, multiple paths and a extent of weight connected with every sensor. Thus, we let, $(i,j)$ indicate the connection from node $i$ to node $j$ for each combination $i,j \in \{1,2 \dots 3\}$, suppose that entire amount beside this link is limited in its capability.

In given model every link is observed as resource and the $(i,j)^{th}$ operation is explained as the procedure of routing packets alongside the comparable link. Objective is to acquire instinct concerning the kind of network load and workload in link-restrained network. With regard to definition of network load, energy dissolution is sufficing to restrain consideration to allocation-estimate vectors. The equilibrium position prescribed through function

$$f(x) = a_0 + \sum_{n=1}^{l} \left( a_n S \frac{n\pi x}{L} + b_n S \frac{n\pi x}{L} \right) \qquad (7.3)$$

Where
$a_0$ is described as the initial amount of energy
$n$ Denotes the no. of nodes
$L$ indicates the number if iterations
$a_n$ Signifies forwarder
$b_n$ Implies transmitters

Assume that $f$ is a maximum flow. Hence none of material is departed in an equilibrium, we requisite to have $f(x) > \alpha *$. So, threshold is preserved for each sensor that accomplishes this lessen bound established on power distribution that obtains the extreme flow $*$.

The set of $A \subset \{1, \ldots l\}$ is described as the set of overall $1 \leq j \leq l$ like that, there occurs a non-binding route from node 1 to node $j$ for the circulation of $f *$. It is supposed that $1 \in A$, and we should have $l \in A^c$ because the flow of $f *$ is supposed to be maximum. AMHRP has the following key attributes are related to the proposed protocol.
- Each network node is fixed in its location.
- Transmission power and range for every sensor node is stable.
- There WBAN Central Controller node (WCC) which act as sink node is attached at the middle of the body(trunk) liable for data gathering of all sensor nodes.
- Sink is supposed to have infinite power resource and extra ordinary capabilities.
- Individual node transmits its data in its particular time slot.
- All nodes in the network primarily know about the location of other sensor nodes.
- Transmission of data beyond sink do not contemplated. Every individual sensor may transmit data through a forwarder or cluster head to destination node i.e. sink node (coordinator).

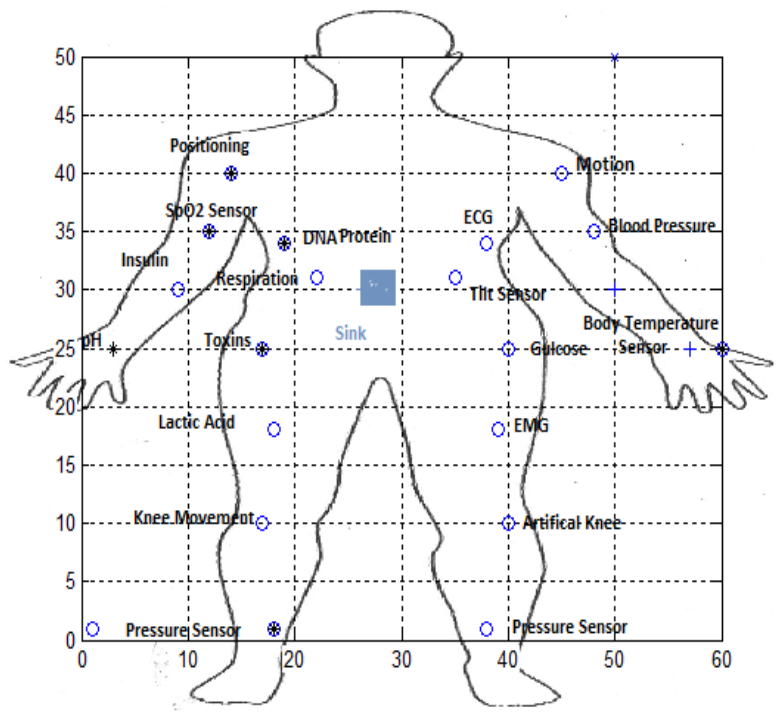

Figure 11: Schematic of Sensor and Sink node on Human body in AMHRP

There are number of parameters to check the performance of a protocol. Performances measuring metrics for AMHRP discussed below:
  i.  **Network Lifetime**

Network lifetime is overall time interval form the formation of network to until death of last node.

####    ii.     Number of alive Nodes
Total amount of sensor nodes which are not exhausted and have quite energy for communication.

####    iii.    Throughput
Throughput is termed as the total no. of data packet successfully delivered to sink in a network.

####    iv.     Residual Energy
Average energy consumption of nodes per round is called residual energy.
####    v.      Path loss:
Decrease in the power of a signal as it broadcast throughout the medium is said to be path loss.it is calculated in decibels (dB).

## 7. Simulation Results and Analysis

For the evaluation of suggested protocol, we conducted a large number of experiments using MATLAB simulator R2012a to analyze the performance of purposed routing protocol. We use 19 sensor nodes which are randomly distributed and sink node is placed in center of network. Total 10000 rounds performed and the initial energy of supposed to 0.5J which can vary. We studied the performance of AMHRP (Adaptive Multi-hop Routing Protocol) and compared it with the existing protocols SIMPLE and M-ATTEMPT.

We stated the transmission with the Poisson threshold value. The diagram shows the network lifetime in worst case where first sensor node dies after completion of 4000 to 5000 rounds. AMHRP in sense of network lifetime performs better as compared to ATTEMPT and has greater lifetime SIMPLE protocol. In M-ATTEMPT, when temperature of forwarder node upsurges, nodes select alternate longer path which consumes more energy. Hence, these nodes die early. Load distribution mechanism and multi-hop communication sustain energy for the node beyond the sink. In ATTEMPT nodes die early due to continuous and single hop transmission. AMHRP achieves 50 % more stability period and 250% longer network lifetime. As shown in figure 12 AMHRP has longer network life time as compared to SIMPLE and ATTEMPT due to more number of nodes remain alive after 10000 rounds.

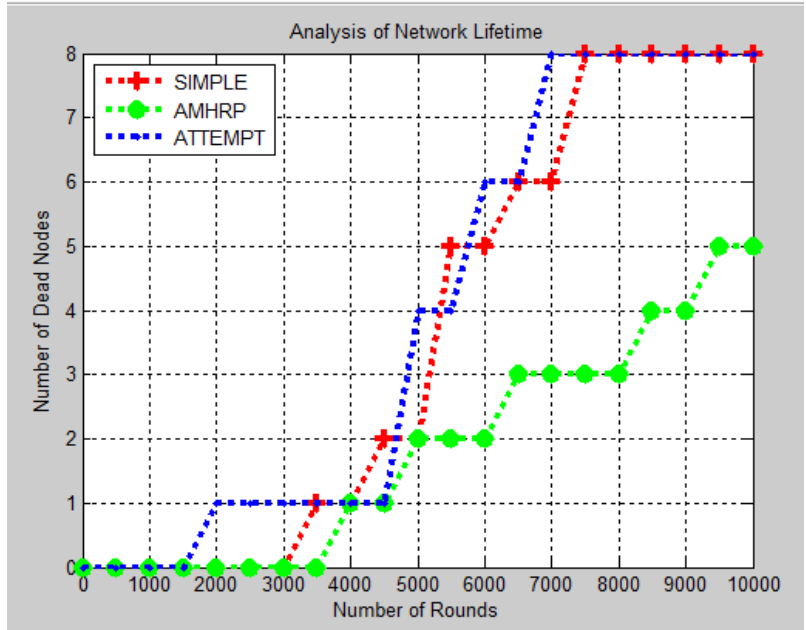

Figure 12: Network Lifetime Analysis

Total number of packets successfully delivered at sink is termed throughput. It is demonstrated as:

$$Throughput(\%) = \frac{Number\ of\ Packets\ Received}{Number\ of\ Packets\ Sent-} * 100 \qquad (7.4)$$

As we know, WBAN has life-threatening and vital data of patient, due to this it needs a protocol which have least packet drop ratio thus gives all-out data received at sink. ATTEMPT dispatches threshold and periodic data, so its throughput is accessed from this data. AMHRP attains high throughput than ATTEMPT and SIMPLE as maximum energy saving and maximum number of nodes stay alive for longer network life as depicted in figure 13. As many nodes remain alive they send max packets to sink which enhance the throughput of network. The stability period of ATTEMP and SIMPLE is smaller than AMHRP it means that they will send small amount of packets to sink. Therefore throughput of ATTEMP and SIMPLE is less than AMHRP. AMHRP have a distributed function so it sends maximum number of packets but energy remains constant.

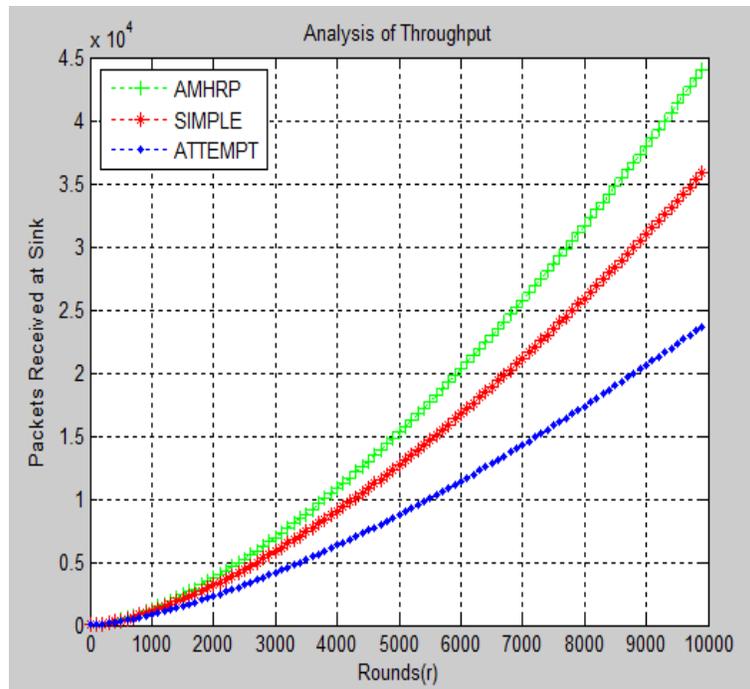

Figure 13 Analysis of Throughput

Figure 14 depicts the normal network energy absorbed in each round. The proposed scheme utilizes multi-hop topology, in which each farthest node transmits its data to sink through a distributed function. To transfer packets to sink, our multi -hop topology employ distributed function and it is strong enough that sending, receiving and forwarding is strong enough that it alive more time than SIMPLE and ATTEMP.

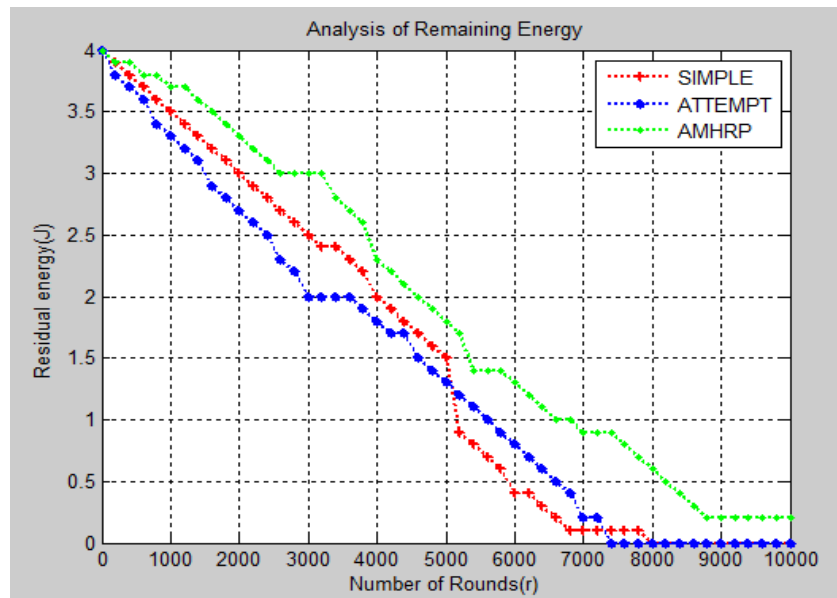

Figure 14: Analysis of Remaining Energy

Simulation outcomes shows that AMHRP absorbs least energy till 100% simulation time as for as for 10000 rounds. It illustrates that, in stability period many nodes have much energy that they can transmit additional data packets to sink till 20000 rounds. In contrast, in ATTEMPT few nodes finishes early because of weighty traffic load and performance of SIMPLE vary during the network period.

Figure 15 illustrated the path loss of various sensor nodes. Path loss is described as the function of distance and frequency. Depicted path loss is function of distance. It is computed from its distance to sink and as computed its distance from sink with persistent frequency 2.4GHz. Aforementioned, multi-hop topology of AMHRP minimize the path loss.

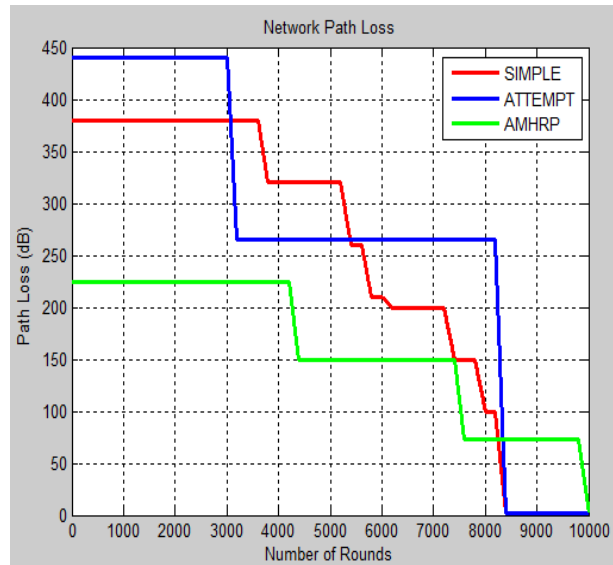

Figure 15: Analysis of Path loss

Reason is that the multi-hop, multipath minimizes the distance due to which path loss decreases. The figure 15 depicts that the AMHRP has the minimum path loss as compared to the simple and attempt because it has a greater number of alive node till the end 8000 node, which lessen the distance to the sink.

The term path loss measured in the form of decibels (dB) and it signifies the signal attenuation. It is notable that the signal power is corrupted by a factor called Additive White Gaussian Noise (AWGN). Path loss is defined as variance amongst the transferred power and the received power of any device however antenna gain isn't mandatory to be considered. The reason of path loss takes place is the growing external area of propagation wave obverse. When transferring antenna discharges its power outward any entity among transmitter and receiver results destruction of originated signal. In field of WBAN, movement of human body, dissimilar body positions, hands, clothes and other accessories disturbs the transferred signal. Path loss associated with distance and frequency is stated as given in [68].

$$PL(f,d) = PL(f) \times PL(d) \qquad (7.5)$$

The relativity of frequency with path loss is stated as

$$\sqrt{PL(f)} \propto f^k \qquad (7.6)$$

In which k is considered as frequency reliant feature and it is interconnected to the geometry of human body. The formulation of distance with path loss is specified here

$$PL(f,d) = PLo + 10n\log_{10}\frac{d}{do} + X\sigma \qquad (7.7)$$

in which PL represents received power, d indicates the space among transmitter and receiver and *do* is considered as referenced distance, in equation value of n relies on propagation environment and it is the path loss coefficient. If there is a free space value of n will be 2, but for WBAN differs from 3 to 4 for line of sight (LOS) type communication and for non- line of sight (NLOS) 5- 7.4. Gaussian random variable is denoted by X and for regular deviation σ is used [69]. PLo indicates the received power at reference distance $d_o$ so it is conveyed as:

$$PL_O = 10\log10\frac{(4\pi \times d \times f)^2}{c} \qquad (7.8)$$

In which f indicates frequency, d is distance (between transmitter and receiver), c for speed of light and the given value of reference distance is 10 cm. In realism it is hard to forecast the power of signal transferred among the edges of transmitter and receiver. For the solution of this question, deviation variable $X\sigma$ is used.

## 8. Conclusion

There is no any certainty, that biological exploration is creating tremendous development at an extraordinary pace. However, meanwhile many sectors are still in inception as the research body is first ever facing such plenty and continued rush of data. Although, data reliability has gained, and its correctness is enhanced, an enormous part of our considerations relies upon the probability of forming the data in a consequential way. Consequently, we can see that new experiments are forefront in which networks and dynamics of networks are essential. Applications gazing from drug design to bacterium host communication to disease network just offset to be analyzed. Bio sensor can play a key role in this regard, by using its characteristics we can explore new prospective in the field of medical health. WBASN technology offers the prototype towards proactive management by concentrating on anticipation and pre-detection of different types of diseases. This can bring revolution in the coming generation health care concerns and decreases the cost of health-related issues. Due to distinctive in-body and on – body constraints designing a routing protocol for WBASNs is a challenging task. The suggested routing protocols either do not take into consideration postural body movements with mobility or are not as energy efficient. Also, many of the above-mentioned routing protocols did not painstaking reliability and QoS.

Our designed protocol with the help of biosensors has gained its objectives as compared to other protocol. The effort bestow in this thesis has the goal to improve the life time of protocol as well as enhance the throughput and residual energy average. Here we descried the conclusion of the thesis and the general prospective of research for future. In this work, a protocol with biosensor with the help of distribution model for observing patients has been proposed. Fix node with Poisson distribution and ensemble random deployment with combination of multi-hop topology is the best combination for the improvement of network life time. To authenticate the correctness of the proposed protocol, simulation have been done using MATLAB. Various parameters have been considered related to routing issues and then make comparison with M-ATTEMPT and SIMPLE. The simulation results obviously express that the network lifetime in the sense of maximum node stay alive (for extra number of rounds e.g. 20000 rounds.) and have 50 % better stability period and 250% longer network lifetime. AMHRP sustain 85%  residual energy till the completion of 10000 rounds in comparison with ATTEMPT and SIMPLE our proposed protocol is 100% better than that. It has gain high throughput as maximum number of nodes stay alive so there is small number of packets drop ratio hence the maximum throughput attained if we relate it with SIMPLE and ATTEMPT. Path loss is not considered in routing, yet if we consider this aspect, AMHRP also have improved it.

In real world there is not any protocol which prompt for the whole network parameters may be fully, hence proposed protocol tried best to conquer some parameters, yet many aspects are still in consideration. In future, routing protocols for WBASNs must provide longer network life through energy efficient schemes. They should provide energy efficient and reliable communication between sensor nodes in real-time and non-real-time application. A reliable communication among heterogeneous bio-medical sensor nodes in real time application is also needed. They must take into consideration the issues of Latency, temperature effects, reliability and power consumption for developing better routing protocol models of WBANs. Different routing challenges belongs to WBASNs are measured in different categories of routing protocols, but still lots of work needs to be done in this regard. Moreover, much accurate and proficient network architecture needs to be developed for best routing structure in WBSNs. As Future work, routing protocols for BANs must be capable of gaining the required QoS as well as preserving well-balanced low power energy consumption. These goals can be gained by mutually construction the Mac layer and routing protocols.